# Study of CeNi$_4$Mn by neutron diffraction

Indu Dhiman, A. Das[*], S.K. Dhar[+], P. Raychaudhuri[+], Surjeet Singh[+&], P. Manfrinetti[#]

Solid State Physics Division, Bhabha Atomic Research Centre, Mumbai 400085, India

[+] Department of Condensed Matter Physics and Materials Science,
Tata Institute of Fundamental Research, Colaba, Mumbai 400005, India

[#]Dipartimento di Chimica, Laboratorio LAMIA-CNR-INFM,
Università di Genova, Via Dodecaneso 31, 16146 Genova, Italy

**Abstract**

We report neutron diffraction measurements on CeNi$_4$Mn, which has recently been identified as a soft ferromagnet (T$_c$ ~ 140 K) with a sizeable spin-transport polarization. Our data show conclusively that the Mn atoms occupy a unique site (4c) in the unit cell, which has the symmetry of the cubic MgCu$_4$Sn-type structure. We infer a moment of 4.6 $\mu_B$ on Mn at 17 K, which is oriented ferromagnetically along the {101} plane. The amplitude of the Mn vibrational motion is found to be larger than that of Ce and Ni atoms at all temperatures, thereby lending support to theoretical prediction of rattling phonon modes in this compound.

**Corresponding Author**

A. Das,
Solid State Physics Division,
Bhabha Atomic Research Centre,
Trombay, Mumbai 400085, India,
Email : adas@barc.gov.in
Fax : +91-22-25505151

[&]Present address: Laboratoire de Physico-Chimie de l'Etat Solide, ICMMO, CNRS, UMR8182, Bât 414, Université Paris-Sud, 91405 Orsay, France

**Introduction**

Recently, CeNi$_4$Mn has been reported to be a soft ferromagnet with a large degree of transport spin polarization at the Fermi level [1]. Specifically, the compound orders ferromagnetically with a Curie temperature of nearly 140 K and a coercive field of < 5 Oe at 5 K. The saturation magnetic moment (4.94 $\mu_B$/f.u.) measured at 2 K is close to the spin-only moment of 5 $\mu_B$ for a Mn$^{2+}$ ion. Point contact Andreev reflection measurements at low temperatures reveal a transport spin-polarization of 66 % making this material potentially important for spintronic applications.

Interestingly, the parent compound CeNi$_5$ crystallizing in the hexagonal CaCu$_5$-type structure is Pauli paramagnetic. Substitution of Ni with Mn in CeNi$_{5-x}$Mn$_x$ leads to a change of structure from hexagonal symmetry to cubic, AuBe$_5$ (C15b-type) for x ≥ 0.9 [2 and references there in]. The interesting structural and magnetic properties of CeNi$_4$Mn have already been the subject of band structure investigations [2-4]. These electronic structure calculations indicate that CeNi$_4$Mn is a semimetal of an unusual nature where the Fermi level in both the spin channels falls inside a deep pseudogap about 0.3 eV wide. Attention has also been drawn to a peculiarity of the crystal structure such that the Mn ions are enclosed in a cage much larger than needed for normal metallic bonding. Thus, low frequency rattling phonon modes have been predicted to occur in this compound [3]. To get further insight into the physical behavior of CeNi$_4$Mn we have undertaken a neutron diffraction study and the results are presented in this communication.

**Experimental Details**

The sample used in the present study was prepared by melting the constituents (99.9 wt.% purity Ce, 99.99 wt.% purity Ni and Mn) on a water cooled copper hearth of a high frequency induction furnace. The molten charge was flipped over each time and re-melted four times to ensure homogenization. It was annealed at 830°C for three weeks. A DTA scan was run to check for the reaction type and formation temperature of this phase (accuracy ± 5 °C in the temperature measurement). The scan showed that $CeNi_4Mn$ decomposes on heating at about 890 °C (likely peritectically), then a second and rather long thermal effect begins at about 960 °C and ends at 1100 °C (where the sample is completely melted). The latter feature reproduces on cooling, while the lower temperature exothermic peak experiences undercooling beginning at about 870 °C. Therefore, annealing at around 840 °C is essential to obtain single phase samples.

X-ray analysis was carried out by powder methods. Powder patterns for precise measurement of the lattice constant were obtained by a Guinier-Stoe camera using the Cu K$\alpha$ radiation ($\lambda$ = 1.5418 Å) and pure Si as an internal standard ($a$ = 5.4308 Å); the Guinier patterns were indexed with the help of the LAZY-PULVERIX program [3-LP] and the lattice parameters determined by least-squares methods. Powder data for Rietveld refinement attempts were collected on a Philips PW 1050/81 diffractometer with Bragg-Brentano geometry and Ni-filtered Cu K$\alpha$ radiation in 2$\theta$ steps of 0.02° and a measuring time of 20s/step; a 10-90° range of 2$\theta$ and 4000 profile points were processed by the DBWS-9411 program, using pseudo-Voigt functions for the peak shape and a $5^{th}$ order polynomial for the background.

The magnetization was measured in Quantum Design SQUID magnetometer. The neutron diffraction measurements were carried out on a Powder Diffractometer at Dhruva reactor, using a wavelength of 1.249Å at selected temperatures between 300 and 17 K. The powdered samples were packed in a cylindrical Vanadium container attached to the cold finger of a closed cycle Helium refrigerator. Rietveld refinement of the profiles was carried out using FULLPROF program [5].

**Results and Discussion**

The powder x-ray diffraction pattern of CeNi$_4$Mn at room temperature was identical to that obtained in the previous work [1]. Rietveld refinement was tried out to refine the cubic structure both on the basis of disordered AuBe$_5$-type and the ordered MgCu$_4$Sn-type. The final refinement parameters were nearly the same for both configurations not distinguishing unambiguously between the two possibilities. However, the parameters obtained suggest that the ordered version may be the actual one: D-W = 1.09, $R_{WP}$ = 2.37, $R_P$ = 1.84, $R_B$ = 3.46 for the disordered configuration and D-W = 1.09, $R_{WP}$ = 2.37, $R_P$ = 1.84 and $R_B$ = 3.33 for the ordered MgCu$_4$Sn-type. The lattice parameter of 6.988(2) Å (Guinier pattern) is in good agreement with the value (6.9873 Å) reported in ref. 1. The ferromagnetic transition temperature and the saturation moment are also in agreement with those reported in ref. 1.

Figure1 shows the neutron diffraction patterns at 300 K and 17 K. The lattice constant, $a$ = 6.988 Å, obtained from the refinement of neutron diffraction pattern recorded at 300K is in agreement with that obtained from x-ray studies. However, this lattice constant differs from the cell constant 6.817 Å used by Voloshina et.al. [4] for the electronic structure calculations. The Reitveld refinement was carried out using the following positions for the atoms: Ce(4a) (0, 0, 0); Ni(16e) ( x, x, x ) and Mn(4c) (0.25, 0.25, 0.25). The ideal positions for MgCu$_4$Sn-type structure, which is an ordered variant of the cubic AuBe$_5$-type, space group $F\bar{4}3m$, are: Ce (0, 0, 0), Ni (0.625, 0.625, 0.625) and Mn (0.25, 0.25, 0.25) [6]. Ni forms corner sharing tetrahedra; analogous to spinel structure; with Ni atom position as free parameter. It is of interest to note that when Ni is exactly at (0.625, 0.625, 0.625) then Ni-Ce and Ni-Mn bonds are of the same length. These positions of atoms were reported by Blazina et al. [7] on MgCu$_4$Sn-type, UNi$_4$M (M= In, Sn, Zn) compounds, inferred from their x-ray diffraction data. However, the position of the Ni in CeNi$_4$Mn is more reliably obtained from the present neutron study. The value of x deviates slightly from 0.625 at all temperatures (Table I). Our data show clearly that Mn atoms replace the Ni atoms preferentially at the 4c site leading to cubic MgCu$_4$Sn type structure. The site occupancies were refined and the deviations from the nominal composition were found to be within 1%, indicating that all the sites are fully occupied. The similarity of the patterns between 300K and 17K shows that the crystal symmetry observed at room temperature is retained on lowering the temperature; the volume of the unit cell is found to decrease linearly with temperature. The results of refinement are tabulated in Table I. No superlattice reflection is observed ruling out the possibility of any antiferromagnetic ordering. On lowering temperature below the ferromagnetic transition temperature, T$_C$ (~140K), enhancement in the intensity of some of the low angle fundamental reflections i.e., (111), (200) and (220) is observed. This implies a ferromagnetic ordering of the spins in the sample in confirmation with previously reported magnetization studies [1] and electronic structure calculations.

CeNi$_5$ is known to be a Pauli paramagnet [8]. In the case of CeNi$_4$Mn we find from the refinement of the diffraction patterns that only Mn carries a magnetic moment. This may also be inferred from the structure factor calculations of some of the strongly magnetic reflections. The equations governing the intensities (I$_{hkl}$) of the initial three reflections with the maximum magnetic contribution are given below.

$$I_{111} \propto (4b_{Ce} + 2ib_{Mn} + (-5.7 + i5.7)b_{Ni})^2$$

$$I_{200} \propto (4b_{Ce} - 4b_{Mn} + 0b_{Ni})^2$$

$$I_{220} \propto (4b_{Ce} + 4b_{Mn} + 0b_{Ni})^2$$

where $b_{Ce}$, $b_{Mn}$, and $b_{Ni}$ denote the scattering lengths for Ce, Mn, and Ni, respectively. The (220) and (200) reflections which show significant enhancement in the intensity below the magnetic ordering temperature do not have any contribution from Ni. In particular, for the (200) reflection, the nuclear contribution is negligibly small as compared to the magnetic contribution. This reflection is nearly absent above $T_C$. Moreover, a reasonably good fit is obtained with moment on the Mn site alone. This is in slight disagreement with the conclusions arrived at by the electronic structure calculations, which predict a combined moment of 1.1-1.2 $\mu_B$ on four Ni atoms and a small moment of -0.2 $\mu_B$ on Ce in addition to approximately 4 $\mu_B$ on Mn. However, the absence of strong magnetic reflections in MgCu$_4$Sn-type structure, where only Ni or Ce contribute while Mn does not, leaves some uncertainty in our conclusions as regards the absence of moment on Ni and Ce. At 17 K the refined magnetic moment is 4.6$\mu_B$/f.u. which is comparable with 4.94 $\mu_B$ obtained from M(H) studies at 5 K. The moment is oriented along {101} plane. The variation of the moment with temperature below $T_C$ is shown in the inset of figure 1. It shows a discontinuous behavior at ~ 80K which is not corroborated by bulk measurements, such as, magnetization and resistivity. This behavior is not understood, yet.

It may be worthwhile here to compare CeNi$_4$Fe with CeNi$_4$Mn. The Fe-compound crystallizes in CaCu$_5$ – type, hexagonal phase (space group P6/mmm) [9]. In this structure Ni occupies both 2c and 3g sites. Mossbäuer studies on isostructrual LaNi$_{5-x}$Fe$_x$ [10] and neutron diffraction studies on LaNi$_4$Co reveal that Fe is distributed over both the sites, although not completely randomly [11]. Magnetization studies indicate that the saturation magnetization of CeNi$_4$Fe is lower (~ 3.3 $\mu_B$) as compared to CeNi$_4$Mn. In the case of CeNi$_4$Fe it is also found that Fe induces moment on the Ni site (~0.9 $\mu_B$/Ni) and the $T_C$ is enhanced to 195K. This may be attributed to the smaller distances between adjacent 3d atoms. A comparison of the distances between various atoms in the two cases is shown in Table II. The Ni-Ce and Ce-Mn bond distances in CeNi$_4$Mn obtained agree well with the values reported by Murugan et al. [2].

Thermal parameters, B, derived from the neutron data for all three atoms Ce, Ni and Mn are shown in table I. The thermal parameter B is related to the amplitude of vibration by the formula B = $8\pi^2$ <u$^2$>, where <u> is the amplitude of the vibration. At all temperatures it is found that the value of B for Mn is higher than that of Ce and Ni. This indicates that Mn has higher amplitude of vibration as compared to the other two. Due to anomalously large distance between Mn and its near neighbour coordination atoms Mazin [3] had noted that the Mn in CeNi$_4$Mn is a rattling ion and he had predicted the likely occurrence of rattling phonon modes. The higher vibrational amplitude observed for Mn corroborate the predicted rattling motion. Similarly, in skutterdites PrFe$_4$Sb$_{12}$ rattling motion of Pr ions leads to a higher thermal parameter for Pr ($B_{iso}^{eq}$ = 1.6 Å$^2$) as compared to Fe (0.4 Å$^2$) and Sb (0.55 Å$^2$) [12]. It would be interesting to carry out further studies using inelastic neutron scattering to investigate in detail the nature of this rattling phonon mode.

**Conclusions**

Neutron diffraction studies at 300 K and at several temperatures below 140 K were carried out on CeNi$_4$Mn. It crystallizes in the MgCu$_4$Sn-type structure which is an ordered variant of the AuBe$_5$ type cubic structure. CeNi$_4$Mn orders ferromagnetically below 140 K and within the limitations imposed by the data as mentioned above, we infer that a magnetic moment is carried only by Mn. The refined magnetic moment at 17 K is 4.6 $\mu_B$, in agreement with the value at 5 K obtained from magnetization studies. The thermal parameter B, which is proportional to the vibrational amplitude is higher for Mn as compared to Ce and Ni at all temperatures. This is in agreement with the theoretical prediction of rattling motion of Mn inside the cage of atoms surrounding it.

**References**

1. S. Singh, G. Sheet, P. Raychaudhari, S. K. Dhar, Appl. Phys. Lett., **88** 022506 (2005).

2. P. Murugan, Abhishek Kumar Singh, G. P. Das, and Yoshiyuki Kawazoe, Cond – mat/0510748vl

3. I. Mazin, Phys. Rev. B, **73** 012415 (2006).

4. Elina N. Voloshina, Yuri. S. Dedkov, Manuel Richter, and Peter Zahn, Phys. Rev. B, **73** 144412 (2006).

5. J. Rodriguez Carvajal, Physica B, **192** 55 (1992)

6. P. Villars and L. D. Calvert, Pearson's Handbook of Crystallographic Data for Intermetallic Phases (American Society for Metals, Metal Park, OH, 1989), Vol 1 – 3.

7. Z. Blazna, A. Drasner and Z. Ban, J. Nucl. Mat., **96** 141 (1981)

8. F. Pourarian, A. T. Pedziwiatr, and W. E. Wallace, J. Appl. Phys., **55** 1987 (1984).

9. F. Pourarian, M. Z. Liu, B. Z. Liu, M. Q. Huang, and W. E. Wallace, J. Solid State Chem, **65** 111 (1986).

10. J. Lamloumi, A. Percheron-Guegan, J. C. Achard, G. Jehanno, and D. Givord, J. Phys. **45** 1643 (1984).

11. E. Gurewitz, H. Pinto, M.P. Draiel, and H. Shaked, J. Phys. F **13** 545 (183).

12. N. P. Butch, W. M. Yuhasz, P. C. Ho, J. R. Jeffries, N. A. Frederick, T. A. Sayles, X. G. Zheng, and M. B. Maple, Phys. Rev. B, **71** 214417 (2005).

**Figure Captions**

**Fig 1**. Neutron diffraction pattern of CeNi$_4$Mn at (a) 300 K and (b) 17 K. The inset shows the variation of the refined magnetic moment with temperature below T$_C$.

**Table Captions**

**Table I**. Structural parameters of CeNi$_4$Mn obtained from the refinement of the neutron diffraction pattern at indicated temperatures.

**Table II**. Bond distances obtained from refinement of powder diffraction pattern at 300 K of CeNi$_4$Mn: The bond distance of CeNi$_4$Fe is calculated from the cell parameter given in ref.9

**Table I**:

|  | 17K | 30K | 50K | 70K | 90K | 110K | 130K | 300K |
|---|---|---|---|---|---|---|---|---|
| a (Å) | 6.9706 (1) | 6.9716 (1) | 6.9725 (1) | 6.9740 (1) | 6.9754 (1) | 6.9757 (1) | 6.9757(1) | 6.9885 (1) |
| V(Å$^3$) | 338.71 (1) | 338.84 (1) | 338.97 (1) | 339.19 (1) | 339.40 (1) | 339.44 (1) | 339.45 (1) | 341.31 (1) |
| M ($\mu_B$/Mn) | 4.6 (2) | 4.2 (2) | 3.6 (2) | 3.1 (2) | 2.2 (2) | 2.4 (2) | 2.2 (2) | 0 |
| $B_{Ce}$ (Å$^2$) | 0.2 (1) | 0.3 (1) | 0.3 (1) | 0.3 (1) | 0.3 (1) | 0.4 (1) | 0.4 (1) | 0.6 (1) |
| $B_{Ni}$ (Å$^2$) | 0.23 (1) | 0.26 (1) | 0.27 (1) | 0.29 (1) | 0.32 (1) | 0.30 (1) | 0.30 (1) | 0.67 (1) |
| $B_{Mn}$ (Å$^2$) | 0.85 (13) | 0.93 (14) | 1.17 (16) | 0.88 (14) | 0.96 (14) | 1.16 (15) | 0.79 (12) | 1.42 (14) |
| $x_{Ni}$ | 0.6240 (2) | 0.6236 (2) | 0.6237 (2) | 0.6234 (2) | 0.6238(2) | 0.6244 (2) | 0.6373 (2) | 0.6244 (2) |

**Table II**. Bond distances obtained from the refinement of powder diffraction pattern at 300 K of CeNi$_4$Mn: The bond distances in CeNi$_4$Fe are calculated from the cell constant given in ref.9

| Bond Distances (Å) | Mn – Mn or Fe – Fe | Ni – Ni | Ni – Mn or Ni – Fe | Ni - Ce | Ce – Mn or Ce – Fe |
|---|---|---|---|---|---|
| CeNi$_4$Mn | 4.94 | 2.46 | 2.89 | 2.90 | 3.02 |
| CeNi$_4$Fe | 1.42 | 2.46 | 2.18 | 3.19 | 3.02 |

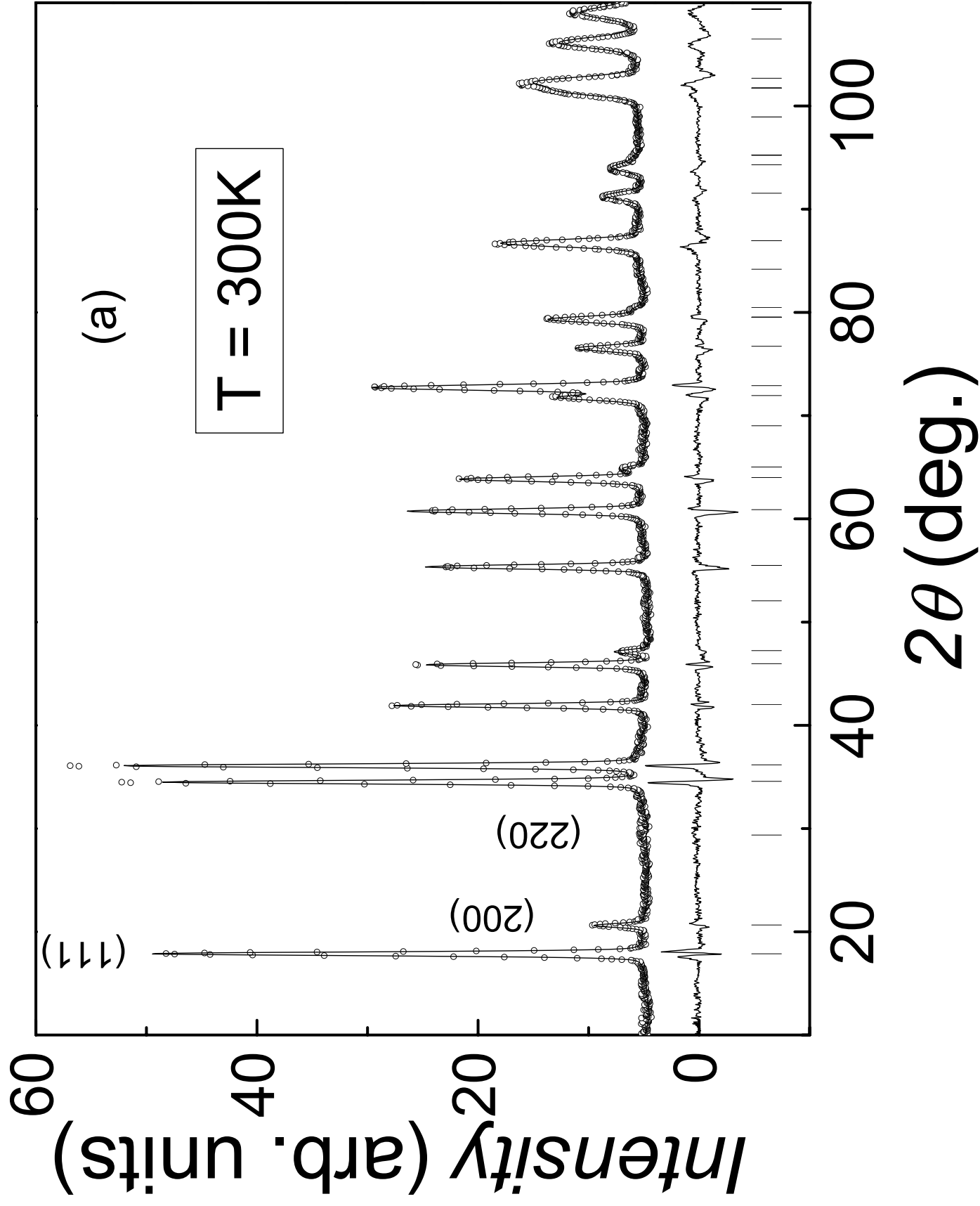

Figure 1a

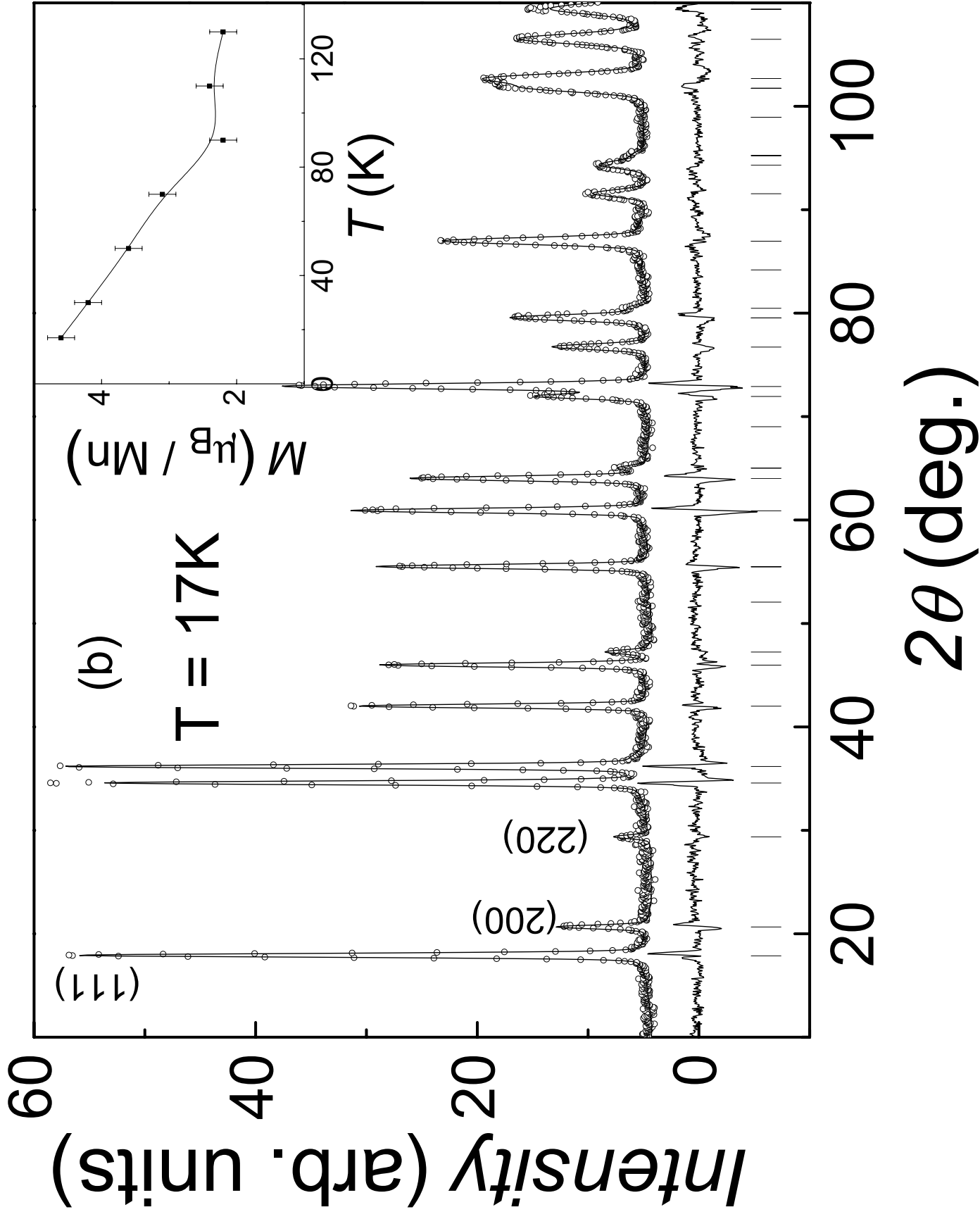

Figure 1b